\documentclass[10pt,conference]{IEEEtran}
\IEEEoverridecommandlockouts
\usepackage{cite}
\usepackage{amsmath,amssymb,amsfonts}
\usepackage{algorithmic}
\usepackage{graphicx}
\usepackage{textcomp}
\usepackage{xcolor}

\IEEEoverridecommandlockouts

\usepackage{subcaption}
\usepackage{amsmath,amsfonts,graphicx}

\usepackage[T1]{fontenc} % add special characters (e.g., umlaute)
\usepackage[utf8]{inputenc} % set utf-8 as default input encoding
\usepackage{wasysym}
\usepackage{graphicx}
\usepackage{color}
\usepackage{wasysym}
\usepackage{graphicx}
\usepackage{textcomp}
\usepackage{xcolor}
\usepackage{tabularx,adjustbox,booktabs}
\usepackage{colortbl}
\usepackage{multirow}
\usepackage{subcaption}
\usepackage{mathtools}
\usepackage{makecell}
\usepackage{lineno}
\usepackage{hyperref}
\nolinenumbers
\definecolor{darkslate}{HTML}{373E40}
\definecolor{tealgreen}{HTML}{488286}

\begin{document}
% ------
\title{MusicLIME: Explainable Multimodal Music Understanding\\}
%

% \name{
% Theodoros Sotirou, Vassilis Lyberatos, Orfeas Menis Mastromichalakis, Giorgos Stamou
% }

% \address{
% National Technical University of Athens, Athens, Greece
% }

\author{
\centering
\IEEEauthorblockN{
Theodoros Sotirou$^{\star}$, 
Vassilis Lyberatos$^{\star}$, 
Orfeas Menis Mastromichalakis$^{\star}$, 
Giorgos Stamou$^{\star}$
}
\IEEEauthorblockA{
$^{\star}$\textit{National Technical University of Athens}, Athens, Greece \\
theodorossotirou@gmail.com, \{vaslyb, menorf\}@ails.ece.ntua.gr, gstam@cs.ntua.gr
}
}

\small
%\ninept
%

\maketitle
\begin{abstract}

% Music understanding tasks, including genre and emotion recognition, are increasingly being tackled using multimodal deep learning models.%\cite{ma2024foundation}
% This trend is driven by the availability of large datasets on the web and advances in algorithms and computational power. However, despite these developments, the reliability and fairness of these models in real-world scenarios remain questionable. The distribution of a model's training data often significantly influences its performance. When this data is biased, the deep learning model's predictions can perpetuate these biases, leading users to follow popular trends in music rather than making independent choices in how they organize and explore music. In this context, interpretability is crucial. It empowers users to responsibly engage with these systems and better understand the rationale behind the model's decisions. Recent interpretable pipelines in music primarily base their explanations on unimodal representations. While these methodologies offer valuable insights into how models operate, they overlook the potential information that may be hidden in other modalities. In this work, we present \textsc{\textsc{MusicLIME}}, a model-agnostic and interpretable pipeline designed for multimodal music understanding.

Multimodal models are critical for music understanding tasks, as they capture the complex interplay between audio and lyrics. However, as these models become more prevalent, the need for explainability grows—understanding how these systems make decisions is vital for ensuring fairness, reducing bias, and fostering trust. In this paper, we introduce \textsc{MusicLIME}, a model-agnostic feature importance explanation method designed for multimodal music models. Unlike traditional unimodal methods, which analyze each modality separately without considering the interaction between them, often leading to incomplete or misleading explanations, \textsc{MusicLIME} reveals how audio and lyrical features interact and contribute to predictions, providing a holistic view of the model's decision-making. Additionally, we enhance local explanations by aggregating them into global explanations, giving users a broader perspective of model behavior. Through this work, we contribute to improving the interpretability of multimodal music models, empowering users to make informed choices, and fostering more equitable, fair, and transparent music understanding systems. 

\end{abstract}
\begin{IEEEkeywords}
Explainable Artificial Intelligence, Music Understanding, Multimodality
\end{IEEEkeywords}

% A notable example is emotion, which is challenging to measure in music due to its subjective and ambiguous nature.
% %~\cite{buisson2022ambiguity}. 
% By leveraging multiple modalities, it is possible to achieve a more comprehensive quantification of emotion, as it can be correlated with both lyrics and audio.

\section{Introduction}\label{sec:intro}

\begin{figure*} 
    \centering
    \includegraphics[width=0.89\textwidth]{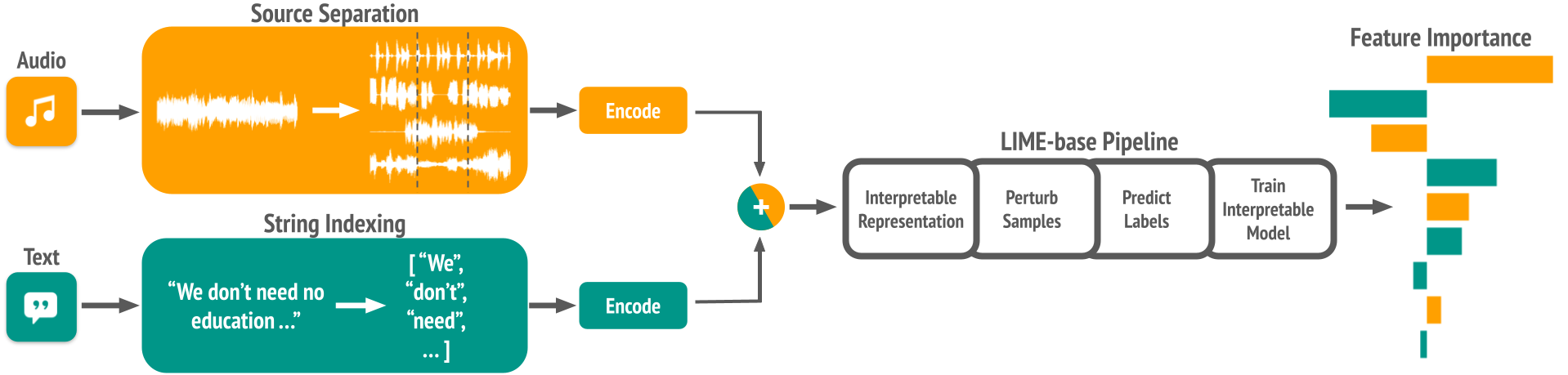}
    \caption{Overview of \textsc{MusicLIME}.}
    \label{fig:multimodal_explanations_pipeline}
\end{figure*}

% While Deep Learning (DL) is pushing the boundaries of Artificial Intelligence (AI), researchers have been increasingly shifting their focus towards the development of multimodal approaches, to leverage the strengths of DL in integrating diverse types of data, enhancing both the complexity and accuracy of AI systems~\cite{han2023survey}. Multimodal models have recently gained traction in the Music Information Retrieval (MIR) domain~\cite{simonetta2019multimodal}. Several studies integrate both audio and lyrical data to build more advanced models for various music analysis tasks, including mood classification~\cite{musicmoodmultimodaldnn}, emotion recognition~\cite{krols2023multimodality}, and music auto-tagging~\cite{huang2022mulan}. Additionally, there is a growing focus on multimodal foundational models in MIR~\cite{ma2024foundation}. The challenge with these pipelines is that, in most cases, the model's workings are not transparent, making it harder to debug and ensure reliable and fair predictions.

As artificial intelligence (AI) continues to evolve, researchers are increasingly focusing on multimodal approaches to harness the strengths of deep learning (DL) across diverse types of data. These multimodal models integrate various data sources, such as text, audio, and images, to enhance accuracy and make better use of the available data~\cite{han2023survey}. In the Music Information Retrieval (MIR) domain, multimodal approaches are becoming increasingly prominent as they combine audio and lyrical data to achieve more precise music analysis~\cite{simonetta2019multimodal}. This includes tasks such as mood classification~\cite{musicmoodmultimodaldnn}, emotion recognition~\cite{krols2023multimodality}, and music auto-tagging~\cite{huang2022mulan}. 
However, the complexity of multimodal models amplifies transparency challenges. The interaction between modalities makes understanding their decisions harder, adding to the transparency issues already present in unimodal systems. This lack of interpretability can obscure the decision-making process, impacting the reliability and fairness of the models.

Explainable AI (XAI) has emerged as a crucial area of research focused on enhancing the interpretability and transparency of machine learning models \cite{mastromichalakis2024rule, liartis2023searching}. XAI methods are essential for understanding how models make decisions and the underlying data \cite{menis2024semantic}, thereby improving user trust and facilitating responsible AI deployment~\cite{panigutti2023co, menisexplainable}. Among these methods, Local Interpretable Model-agnostic Explanations (\textsc{LIME}) stands out as a seminal and widely accepted approach in the XAI field~\cite{lime}. It provides local explanations by systematically perturbing input features and observing how predictions change, offering a valuable tool for examining model behavior at the instance level. Recent advances in the area include \textsc{AudioLIME}, a variant of \textsc{LIME} adapted specifically for the audio domain, which applies the same principle to audio-specific features~\cite{audiolime}. In the music domain, XAI methods have been applied to interpret models through attention mechanisms~\cite{selfattentionmusicinterpretability}, perceptual musical features~\cite{lyberatos2024perceptual}, genre and spectral prototypes~\cite{alonso2024leveraging, loiseau22amodelyoucanhear}, and concept-based explanations~\cite{foscarin2022concept}. 
% However, while these efforts have advanced XAI in music, a notable gap remains in developing explainability methods for multimodal models that integrate both audio and lyrical data.

% While several studies have explored the development of XAI methodologies for multimodal settings~\cite{multimodalexplainabilityreview}, 

% \textsc{LIME} has also been applied to multimodal data, in \cite{al2023interpretable} they developed a multimodal \textsc{LIME} for image and text in order to build a sentiment classification model. 

% none have been applied to MIR tasks. In this paper, we introduce \textsc{MusicLIME}, a model-agnostic explainable approach designed specifically for music understanding tasks, with a focus on models that integrate both audio and text modalities.
While existing XAI methods have advanced explainability in the music domain, there is a notable gap in approaches tailored to multimodal models, particularly in music, which combines both audio and lyrical data. Multimodal explainability offers a significant advantage over unimodal methods by providing a more comprehensive understanding of how different modalities interact within a model's decision-making process. Unlike unimodal approaches, which analyze each modality in isolation and can lead to incomplete or misleading insights, multimodal explanations enable a holistic overview of the model’s behavior by revealing the contributions and interactions between features from different modalities.  This allows users to better understand the intricate dynamics between, for example, lyrical content and audio features in music. Several studies have explored XAI methodologies for multimodal settings~\cite{multimodalexplainabilityreview}, including the development of a multimodal \textsc{LIME} approach for image and text sentiment classification~\cite{al2023interpretable}. However, these methodologies have yet to be fully applied to the MIR domain, leaving a gap in explainability for multimodal music models.

% Our contributions include curating two datasets specifically tailored for multimodal music emotion and genre recognition. We developed an advanced multimodal model leveraging transformer-based architectures to address these tasks. Furthermore, we introduce a novel methodology for enhancing multimodal explainability in music. Finally, we analyze the performance of our models and discuss potential biases in both the data and models that could impact the results.

In this paper, we introduce \textsc{MusicLIME}, a model-agnostic feature importance explanation method specifically designed for multimodal music understanding systems. As part of our methodology, we curated two datasets tailoring them for multimodal music emotion and genre recognition and developed a transformer-based multimodal model to tackle these challenges.  \textsc{MusicLIME} addresses the challenge of explaining the interactions between audio and lyrical features, providing a more comprehensive view of how these modalities contribute to predictions. Additionally, we provide global explanations by aggregating local explanations, offering a broader understanding of the model’s overall behavior. All code, implementation details, and instructions for reproducing the results are available in our GitHub repository~\footnote{https://github.com/IamTheo2000/MusicLIME}.

\section{Methodology}

% Single Modality (just references/intro)
% \subsection{Model architectures}
% After a thorough investigation of feature engineering and model architectures, we selected RoBERTa \cite{roberta} and AST \cite{ast} to process lyrics and audio, respectively. To evaluate model performance and behavior for each classification task, we employed three models: one for each of the lyrical and audio modalities, and a third combinational model. The lyrical model harnesses the power of the pretrained RoBERTa-large architecture for text classification by first preprocessing and tokenizing the input text. The audio model utilizes the pretrained Audio Spectrogram Transformer for audio classification, loaded from a specifically finetuned checkpoint. We used the provided Feature Extractor to generate mel-spectrograms and further finetuned the model. Finally, our multimodal approach integrates the previous architectures by concatenating the pooled output from AST with the CLS token from RoBERTa. This unified feature vector is then processed by a classification head, consisting of a normalization layer and a fully connected layer.

\subsection{Model Architecture}

% In our study, we conducted experiments using two different modalities: text and audio. For model construction, we integrated a language model, a transformer-based model for audio, and a model that combines both modalities by concatenating their embeddings into a single vector and feeding it into a classification head. Our goal is to evaluate overall model performance and, more importantly, explore and compare the behaviors of unimodal and multimodal architectures across the tasks of music genre and emotion classification.

We experimented with two modalities: text (lyrics) and audio, utilizing a language model for text and an audio model respectively. These two transformer-based models were combined into a single multimodal model by concatenating their embeddings into a unified vector, which is then fed into a classification head. The aim was to establish a baseline model that will be used to evaluate the effectiveness of our \textsc{MusicLIME} method in providing insights into the model’s behavior across music genre and emotion classification tasks. Notably, our approach can be effortlessly adapted to any model of choice.

After a thorough investigation of model architectures, we choose to experiment with \textsc{RoBERTa} \cite{roberta} as our language model and Audio Spectrogram Transformer (\textsc{AST}) \cite{ast} as our audio model. These models were chosen for their ease of implementation and their balanced size and performance. It is important to note that our methodology is model-agnostic and can be easily applied to larger models. We utilized \textsc{RoBERTa-large}~\footnote{https://huggingface.co/docs/transformers/en/model\_doc/roberta}, by first preprocessing and tokenizing the input lyrics. We generated mel-spectrograms from the audios, using the provided \textsc{Feature Extractor}, specific for the \textsc{AST}~\footnote{https://huggingface.co/docs/transformers/en/model\_doc/audio-spectrogram-transformer}. Our multimodal framework was created by concatenating the pooled output from the \textsc{AST} with the CLS token from \textsc{RoBERTa}. This combined feature vector was then fed into a classification head, comprising a normalization layer and a fully connected layer, to generate the final predictions. Fine-tuning on our dataset was performed both separately for each modality and jointly for the multimodal setting.

% In our study we integrate a language model, a model and one that combines both these modalities. Our goal is to evaluate and overall model performance and, more importantly, explore and compare the behaviors of the unimodal and multimodal architectures across the tasks of music genre and emotion classification. The multimodal architecture combines the language and audio models by concatenating their embeddings into a single vector and feeding it into a classification head.

% Multimodal (write details)
% Orfeas suggestion: This section should have 2 paragraphs as it has now, the first one introducing the unimodal approaches and the second one introducing our multimodal MusicLIME. The separation is basically as it is, but we need to be less "babbler" in the 1st paragraph, and in the 2nd we will start by citing figure 1, and based on that explain our approach.   
\subsection{Unimodal and Multimodal Explainability}
% To gain a clear understanding of how our models make decisions and interpret the data, we adopted the \textsc{LIME} methodology, which provides explanations for individual instances. It is particularly well-suited for our case because it is model-agnostic, provides human-readable explanations, and can be applied to different kinds of data. To approximate the language model's behavior locally, we employed \textsc{LIME} for text, which assigns weights to the words of the input based on their contribution to the model's predictions. In the audio domain, while it is possible to treat spectrograms as images and use \textsc{LIME} for images to highlight the important parts for each instance, the explanations provided are neither understandable nor interpretable. To address this issue, we opted for \textsc{AudioLIME}, which provides intuitive and easily interpretable, listenable explanations. It employs temporal segmentation techniques to divide the audio into distinct time segments, followed by source separation to isolate components such as vocals, drums, bass, and other instruments within each segment. This method enables listeners to clearly identify and focus on the most important elements within an audio segment. 
In this study, we selected \textsc{LIME} as the foundation for our explainability approach due to its simplicity,  widespread adoption, and proven effectiveness in providing intuitive model explanations. \textsc{LIME} has been successfully adapted to various domains and modalities, including images, audio, and text. In the music domain, the two primary modalities of interest are text and audio. For text-based models, \textsc{LIME} assigns importance scores to individual words, indicating their contribution to the final prediction. In the audio domain, while spectrograms can be treated as images to highlight important parts using \textsc{LIME}, such explanations are often difficult to understand or interpret. A more suitable approach is \textsc{AudioLIME}, a specialized version that segments audio into meaningful time intervals and isolates components like vocals or instruments, resulting in more comprehensible and intuitive explanations.

While the aforementioned approaches provide useful explanations for unimodal models, the multimodal nature of music requires an adaptation that can capture the intricate interplay between its different modalities. To address this limitation, we created \textsc{MusicLIME}, an extension of the \textsc{LIME} \textsc{framework} specifically tailored for multimodal music understanding models. \textsc{MusicLIME} is designed to explain how both audio and lyrical features interact and contribute to a model’s predictions, offering a more comprehensive view of the decision-making process in music classification tasks.  An overview of \textsc{MusicLIME} is shown in Figure~\ref{fig:multimodal_explanations_pipeline}.
\textsc{MusicLIME} processes the two modalities separately before integrating them.
%for a comprehensive explanation. 
For the audio modality, our approach builds on \textsc{AudioLIME} by splitting the input into temporal segments and further decomposing each segment into its constituent sources. Additionally, our GitHub repository~\footnote{https://github.com/IamTheo2000/MusicLIME} offers the option to use Demucs, another highly regarded state-of-the-art source separation model~\cite{rouard2022hybrid}.
% However, we enhance the source separation process by utilizing the open-unmix model (UMX)~\cite{umx}, which provides improved separation quality compared to the original \textsc{AudioLIME} method measured in the \textsl{MUSDB18} \cite{MUSDB18} dataset. 
Each audio instance is divided into 10 segments and split into the components: vocals, drums, bass, and other instruments. For the text modality, we follow an approach similar to traditional \textsc{LIME} for text, where the input is split into individual words. After pre-processing, the features from both modalities are encoded and concatenated into a unified feature vector, indicating the presence or absence of features.
Following \textsc{LIME}’s methodology, we perturb this vector representation by selectively including or excluding features, allowing us to observe changes in the model’s predictions. Using these results, we train an interpretable linear model to approximate the multimodal model's behavior locally. This approach enables us to compute feature importance scores for both audio and text simultaneously, facilitating a direct comparison of their contributions to the model's decision-making process.

\subsection{Global Aggregations of Local Explanations}
To gain a comprehensive understanding of the model's behavior beyond individual instances, generating class-wide explanations, we implemented Global Aggregations of Local Explanations as described in \cite{global_aggregations_of_local_explanations}. In our work, we apply two methods: (1) Global Average Importance, and (2) Global Homogeneity-Weighted Importance.

% The Global \textsc{LIME} Importance for a class \( c \) and a feature \( j \) is given by:
% \begin{equation}
%    I_{cj}^{\text{LIME}} = \sqrt{\sum_{i \in S_c} |W_{ij}|} 
% \end{equation}

% where \( S_c \) includes all instances classified as class \( c \), and \( W_{ij} \) is the weight of feature \( j \) for instance \( i \). This metric assumes that features occurring more frequently have a larger impact on model predictions, which may bias the importance towards common features.

% To address this, the Global Average Class Importance adjusts for the frequency of features and is calculated as:
% \begin{equation}
%   I_{cj}^{\text{AVG}} = \frac{\sum_{i \in S_c} |W_{ij}|}{\sum_{i \in S_c: W_{ij} \neq 0} 1}  
% \end{equation}

% The final method involves calculating a normalization vector for each feature \( j \) across all classes:
% \begin{equation}
%   p_{cj} = \frac{\sqrt{\sum_{i \in S_c} |W_{ij}|}}{\sum_{c \in L} \sqrt{\sum_{i \in S_c} |W_{ij}|}}  
% \end{equation}

% where \( L \) is the set of all labels. The normalized \textsc{LIME} importance \( p_j \) represents the distribution of feature \( j \)’s importance across classes. The Shannon entropy of this distribution is calculated as:
% \begin{equation}
%     H_j = -\sum_{c \in L} p_{cj} \log(p_{cj})
% \end{equation}

% The Global \textsc{LIME} Importance for class \( c \) and feature \( j \) is calculated as \(I_{cj}^{\text{LIME}} = \sqrt{\sum_{i \in S_c} |W_{ij}|} \) where \( S_c \) includes all instances of class \( c \), and \( W_{ij} \) is the weight of feature \( j \) for instance \( i \). To address feature frequency bias, 

The Global Average Class Importance is calculated as follows:
\begin{equation}
 I_{cj}^{\text{AVG}} = \frac{\sum_{i \in S_c} |W_{ij}|}{\sum_{i \in S_c: W_{ij} \neq 0} 1}
  \label{eq:average_importance_equation}
\end{equation}
% \(I_{cj}^{\text{AVG}} = \frac{\sum_{i \in S_c} |W_{ij}|}{\sum_{i \in S_c: W_{ij} \neq 0} 1}\)
where \( S_c \) is the set of all instances classified as class \( c \), and \( W_{ij} \) is the weight of feature \( j \) for instance \( i \) provided by \textsc{LIME}.

The second method involves calculating a normalization vector for each feature \( j \) across all classes \( L \) as \( p_{cj} = \frac{\sqrt{\sum_{i \in S_c} |W_{ij}|}}{\sum_{c \in L} \sqrt{\sum_{i \in S_c} |W_{ij}|}} \). The normalized \textsc{LIME} importance \( p_{cj} \) represents the distribution of feature \( j \)’s importance across classes. The Shannon entropy of this distribution is calculated as \( H_j = -\sum_{c \in L} p_{cj} \log(p_{cj}) \), measuring the homogeneity of feature importance across multiple classes. Finally, the Homogeneity-Weighted Importance is:
\begin{equation}
    I_{cj}^H = \left(1 - \frac{H_j - H_{\text{min}}}{H_{\text{max}} - H_{\text{min}}}\right) \sqrt{\sum_{i \in S_c} |W_{ij}|}
    \label{eq:homogeneity_importance_equation}
\end{equation}
where \(H_{\text{min}} \) and \( H_{\text{max}} \) are the minimum and maximum entropy values across all features. Intuitively, this method penalizes features that influence multiple classes, whereas higher weights are assigned to features that are important for specific classes.

Implementing \eqref{eq:average_importance_equation} and \eqref{eq:homogeneity_importance_equation}, we note that for multimodal models, homogeneity-weighted importance does not accurately capture the influence of multimodal features. This is due to the different nature of the features. While words are distinct, audio features encapsulate different sounds. For example, a vocal feature can contain various styles ranging from soothing singing to screams and shouts. As a result, the same audio features can impact many classes for different reasons. Since Homogeneity-weighted importance punishes features that impact multiple classes, lower weights are assigned to audio features compared to the text ones, which is inaccurate. Therefore, global average class importance is more suited for multimodal analysis.

\section{Experiments}

% Dataset
\subsection{Datasets}
% Although the MIR community has created various multimodal datasets~\cite{christodouloumultimodal}, many of which could also be found on ISMIR's resource page\footnote{https://ismir.net/resources/datasets/}, we encountered significant challenges in finding a dataset that includes both audio and lyrics,  primarily due to copyright restrictions.  For this work, we curated two datasets: \textsl{Music4All}~\cite{santana2020music4all} (\textsl{M4A}) and a subset of \textsc{AudioSet}~\cite{audioset}.
Although the Music Information Retrieval (MIR) community has created various multimodal datasets~\cite{christodouloumultimodal}, many of which can be found on ISMIR's resource page\footnote{https://ismir.net/resources/datasets/}, finding a dataset that includes both audio and lyrics remains challenging due to copyright restrictions. For this study, we curated two datasets: \textsl{Music4All}\cite{santana2020music4all} (\textsl{M4A}), a multimodal dataset with both audio and lyrics, and a manually constructed multimodal subset of \textsc{AudioSet}\cite{audioset}, where we combined audio from \textsc{AudioSet} with lyrics sourced from external databases.

\textsl{M4A} provides 30-second audio clips and lyrics for each instance, along with metadata including genre labels and valence-energy values. Using these metadata, we categorized the songs into nine distinct genres (\textit{heavy music}, \textit{punk}, \textit{rock}, \textit{alternative rock}, \textit{folk}, \textit{pop}, \textit{rhythm music}, \textit{hip hop} and \textit{electronic}) based on Musicmap \footnote{https://musicmap.info/} and nine distinct emotion categories derived from Russel's circumplex model \cite{emotionclassificaitonrussel}. Songs that did not fit into one of these nine categories, such as soundtrack music, were excluded from the final dataset. 
% To reduce the presence of noisy labels, we compared the genre labels from \textsl{Music4All} with Spotify's artist genres. 
% Although we tried to address the imbalance in the emotion labels by doubling the population of the least populated class, this class remains underrepresented. 
% The final refined dataset contains approximately 60,000 songs, 50,000 of which were used for training. We ensured that in the train-validation split the same artist did not appear in the test split to evaluate the model's performance on truly unseen data and ensure its ability to generalize to new, unheard artists.
To ensure label accuracy, we cross-referenced the genre labels with Spotify’s artist genre classifications, refining the dataset to include around 60,000 songs, with 50,000 reserved for training. We maintained a data split where no artist from the training and validation sets appeared in the test set, ensuring that the model was evaluated on truly unseen data for generalizability.

% We tailored \textsl{AudioSet} to further test our models, generate additional explanations, and prove that our results are not dataset-dependent. This dataset includes millions of audio recordings from YouTube with corresponding labels, serving as descriptions of the audio (e.g. \textit{fireworks} or \textit{harmonica}). We focused on the music recordings and selected those matching the genre labels required for our study. 
% Augmenting this dataset with lyrics was difficult, involving fetching video titles, filtering out non-song instances (such as compilations, remixes, and series episodes), extracting artist and song names from the titles, and then retrieving the corresponding lyrics if available. 
% Despite its large scale, incorporating lyrics constrained the number of usable recordings to only 308. This process involved fetching video titles for all entries, filtering out non-song instances (compilations, remixes or even series episodes), extracting artist and song names from the titles and finally retrieving the corresponding lyrics from two sources \footnote{https://docs.genius.com/} \footnote{https://lyrics.lyricfind.com/} if available. 

To further validate our methodology and ensure that our results are not dependent on a single dataset, we created a small multimodal dataset based on a subset of music-related recordings from \textsc{AudioSet}~\cite{audioset}. \textsc{AudioSet} contains descriptive labels (e.g., \textit{fireworks}, \textit{harmonica}) of YouTube videos' audio segments. We focused on music samples and matched the song titles with lyrics from two openly available sources\footnote{https://docs.genius.com/}\footnote{https://lyrics.lyricfind.com/}. This process involved fetching video titles for all entries, filtering out non-song instances (such as compilations, remixes, or series episodes), extracting artist and song names from the titles, and retrieving the corresponding lyrics when available. This procedure resulted in a set of 308 audio-lyrics pairs, which were used to evaluate the robustness of our approach across different datasets, thereby introducing a new curated multimodal music dataset.

%, though we had to merge the rock, alternative rock, and punk categories due to the limited population.

% Experimental setup
\subsection{Experimental Setup}
Our configurations utilized NVIDIA's V100 and P100 GPUs, with 16 GB of RAM each. All models were implemented using the PyTorch framework, with additional utility libraries provided by Hugging Face. A preprocessing step was necessary for our data. For the textual data, this involved standard data-cleaning procedures, such as converting characters to lowercase and removing punctuation. After cleaning, the text was tokenized into sequences of up to 256 tokens. For the audio data, we extracted mel-spectrograms with 128 mel bands, utilizing a window and FFT size of 512, with a sampling rate of 44100 Hz. For the training of each model, default values for learning rates, batch sizes, and the number of epochs were utilized. A checkpointing mechanism was implemented throughout the training process to ensure that the model state corresponding to the highest validation accuracy was preserved.

% This procedure resulted in an input MFCC \begin{math} K \in \mathbb{R} ^{S \times P} \end{math}, where $S$ is the number of segments and $P$ is the number of MFCCs.

% \begin{table}
% \centering
% \small % You can also try \footnotesize or \scriptsize
% \begin{tabular}{
%   >{\columncolor{tealgreen}\color{white}}c 
%   >{\columncolor{tealgreen}\color{white}}c
%   >{\columncolor{tealgreen}\color{white}}c
%   >{\columncolor{tealgreen}\color{white}}c
%   >{\columncolor{tealgreen}\color{white}}c
% }
% \rowcolor{darkslate}
% \textcolor{white}{Model} & \textcolor{white}{lr} & \textcolor{white}{Optimizer} & \textcolor{white}{Batch Size} & \textcolor{white}{Epochs} \\ 
% RoBERTa            & 9e-7  & AdamW  & 10  & 9 \\
% AST                & 6e-6  & Adam   & 8   & 5 \\
% RoBERTa + AST      & 3e-6  & Adam   & 3   & 6 \\
% \end{tabular}
% \caption{Hyperparameters for Different Models}
% \label{tab:model_hyperparameters}
% \end{table}

\begin{figure*}[!h]
    \centering
    \includegraphics[width=0.85\textwidth]{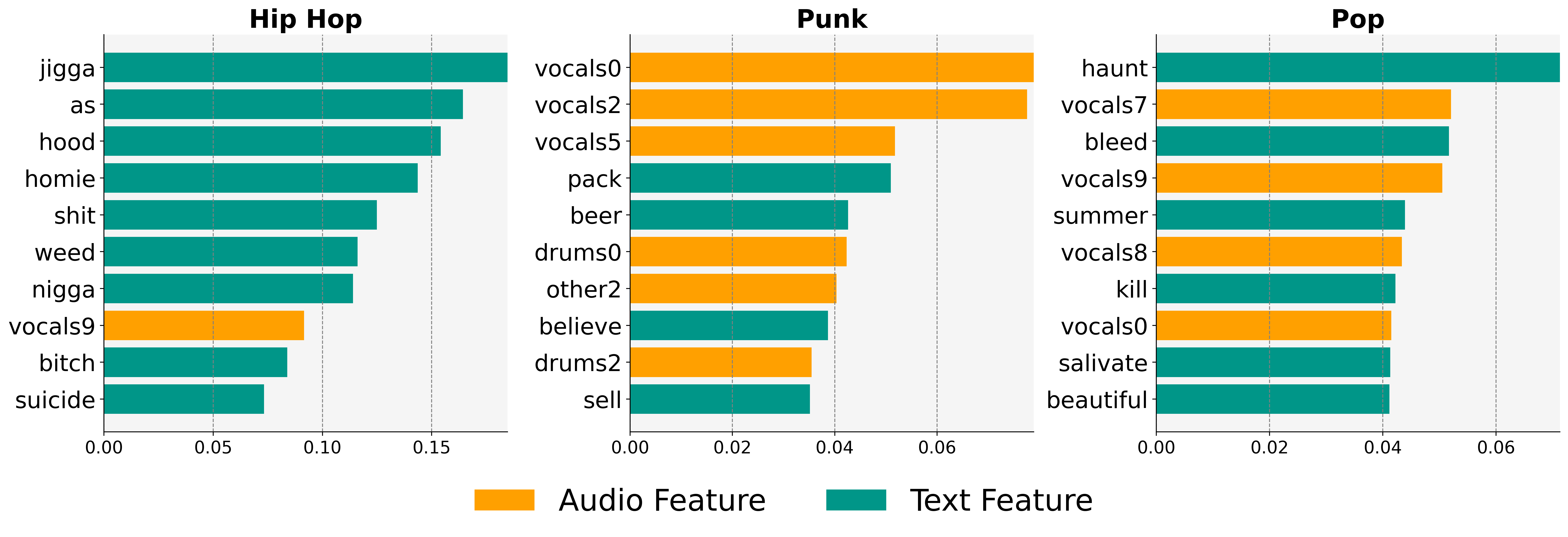}
    \caption{Top 10 features from the global aggregates for the \textit{hip hop}, \textit{punk}, and \textit{pop} genres from the \textsl{Music4All} dataset.}
    \label{fig:multimodal}
\end{figure*}

% , only around 10.2 seconds of the provided audio was used to create the spectrograms.

To generate the global aggregates, we combined the weights produced by multiple instances, each generated with a different number of perturbations. Specifically, for the \textsl{M4A} dataset, we aggregated the results from 640 instances for the lyrical model, 240 instances for the audio model, and 128 instances for the multimodal model. For the \textsl{AudioSet} dataset, we combined the results of all the instances for the language model and 232 for the audio and multimodal models. The number of perturbations per instance for the language, audio, and multimodal models were 2500, 2000, and 5000 respectively. Finally, to visualize the aggregate weights of the words for each class and facilitate comparisons, we employed GloVe embeddings combined with \textsc{t-SNE} for dimensionality reduction. 
% \textsc{K-means} clustering was employed to help identify significant themes influencing the model's decisions (see Figure~\ref{fig:datsets_vs}). 
% For the audio modality, heatmaps were generated to visualize feature weights across temporal segments and audio sources.
\begin{figure}[!b]
    \centering
    \includegraphics[width=0.49\textwidth]{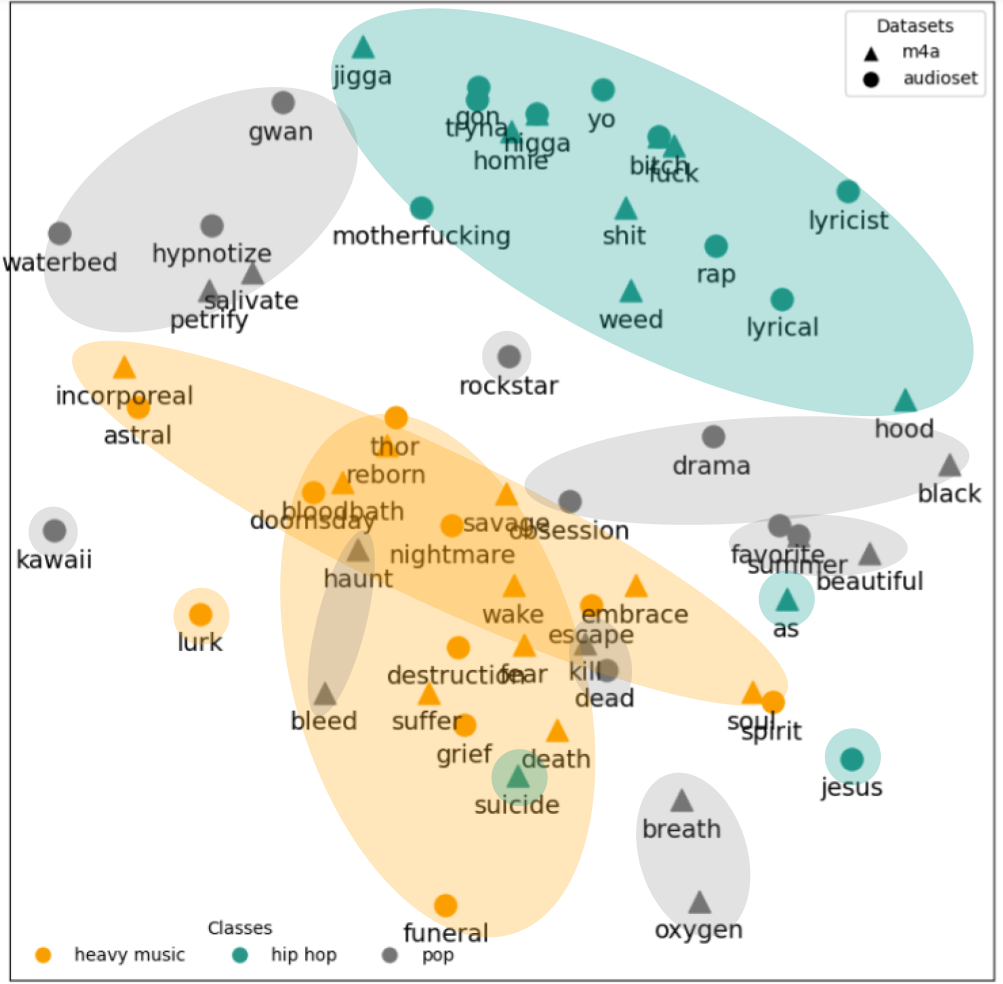}
    \caption{Top 10 lyrical features for the \textit{heavy music}, \textit{hip hop}, and \textit{pop} genres for both datasets clustered.}
    \label{fig:lyrics}
\end{figure}
\section{Results \& Discussion}

\begin{table}[t]
\centering
% \small % You can also try \footnotesize or \scriptsize
\begin{tabular}{l c c}
\hline
\textbf{Model}  & \textbf{Test Acc.} \\ 
\hline
Lyrical Emotion (RoBERTa)&  32.33\% \\
Audio Emotion (AST)    &  48.29\% \\
Multimodal Emotion  & \textbf{48.53}\% \\
\hline
Lyrical Genre (RoBERTa)&  45.14\% \\
Audio Genre (AST)&  53.75\% \\
Multimodal Genre    &  \textbf{57.34}\% \\
\hline
\end{tabular}
\caption{Model performance summary.}
\label{tab:model_performance}
\end{table}

Table~\ref{tab:model_performance} summarizes the performance of our models on the \textsl{M4A} dataset. Overall, the multimodal model consistently outperforms the unimodal models, demonstrating the value of combining text and audio features in music classification. The language model showed limited accuracy in predicting emotions but performed really well at identifying specific genres, such as \textit{hip hop} and \textit{heavy music}, likely due to recurring thematic elements in the lyrics, as further elucidated by our explanations (see Figures~\ref{fig:multimodal} and \ref{fig:lyrics}). Conversely, the audio model, generally outperformed the lyrical model across tasks, especially in emotion classification, indicating that mood-related information is more effectively captured in the audio domain. Additionally, genre prediction proved more accurate than emotion prediction, which may be attributed to the inherently subjective nature of human emotions~\cite{emotionlabeling} on one side, but also to the distinctive features of various genres, whether in lyrics (e.g., hip hop) or audio (e.g., vocals and drums in punk music). These observations are further validated by our multimodal explanations presented in the following paragraphs. 
% and the difficulty of providing as well as detecting a generalized, objective, and therefore accurate estimates of the emotional characteristics listeners might perceive in a track. 
Overall, the results emphasize the complementary strengths of each modality and highlight the importance of using multimodal explanations to better understand model behavior.

% multimodality xai

Figure~\ref{fig:multimodal} demonstrates the effectiveness of our approach and highlights its advantages over unimodal explanations. The Figure presents global multimodal explanations for \textit{hip hop}, \textit{punk}, and \textit{pop}, with teal (greenish) representing lyrical features and amber representing audio features. For \textit{hip hop}, the explanations reveal that lyrical features predominantly drive the model's decision. In contrast, for \textit{punk} music, audio features appear to play a more significant role. For \textit{pop} music, neither audio nor lyrical features dominate, suggesting a more balanced influence from both modalities. These insights, which cannot be fully derived from unimodal explanations due to the lack of direct comparison between feature importances, align with the nature of each genre. \textit{hip hop}'s strong lyrical focus, \textit{punk}’s distinctive musical characteristics, and \textit{pop}'s more diverse thematic content are reflected in the explanations. These findings are further supported by the global lyrical explanations shown in Figure~\ref {fig:lyrics}.  This 2D visualization of the top 10 lyrical features for \textit{hip hop}, \textit{heavy music}, and \textit{pop} reveals that genres where lyrical features dominate also exhibit distinct thematic topics. For instance, \textit{hip hop} features predominantly revolve around street culture, slurs, slang, and artistic expressions, while \textit{heavy music}'s lyrical content centers on dark themes and fantasy elements. Conversely, \textit{pop} music’s lyrical content lacks a dominant thematic focus, leading the multimodal model to rely on both audio and lyrics for accurate genre classification.

%multimodality music

Our findings, further detailed on our GitHub repository~\footnote{https://github.com/IamTheo2000/MusicLIME}, align with the established characteristics of various music genres and associated emotions, supporting the robustness of our methodology~\cite{worlu2017predicting}. The multimodal explanations we identified align with the anticipated genre-specific and emotion-specific features. For instance, \textit{folk} music frequently incorporates regional instruments and lyrics that reflect rural life. In contrast, \textit{electronic} music is characterized by the prominence of drums and synthesizers. Similarly, the presence of guitars is a defining feature in \textit{rock} music. Regarding emotion tags, the \textit{tense} emotion appears to be strongly associated with vocal elements, likely due to its connection with the \textit{hip hop} genre. Additionally, positive emotions such as \textit{happy} and \textit{exciting} are often correlated with the use of drums, possibly due to their powerful and dynamic sound.

It is noteworthy that the multimodal explanations produced by \textsc{MusicLIME} are consistent with the observations and assumptions that a user makes based on the performance metrics outlined in Table~\ref{tab:model_performance}. In music emotion recognition, audio emerges as the dominant modality, as evidenced by the marginal performance improvement when incorporating multimodal inputs and the predominance of audio-based features in the explanations for emotion predictions. This result is in strong agreement with the relevant literature~\cite{rachman2020music,ali2006songs,xu2021using}. Conversely, in genre recognition, both modalities contribute significantly, enhancing overall model performance and yielding explanations that attribute nearly equal importance to each modality.

\section{Conclusions \& Future Work}

In this study, we investigated deep-learning-based multimodal models, curated two multimodal music datasets, and introduced \textsc{MusicLIME}, a novel, model-agnostic explanation methodology specifically designed for music understanding. Our findings highlight that multimodal approaches outperform unimodal ones by leveraging the complementary information embedded in different modalities and their interactions. We also developed a global aggregation methodology that enhances the interpretation of the relationships between genres or emotions and their associated audio and lyrical features, providing a comprehensive view of the most representative characteristics of each class. We assessed the robustness of \textsc{MusicLIME} through its application to two distinct datasets and tasks, demonstrating its effectiveness in various contexts.

% In this study, we examined deep-learning multimodal models and how to gain insights into their workings. We introduced \textsc{MusicLIME} a novel multimodal, model-agnostic explanation methodology specifically designed for music understanding. Our findings demonstrate that multimodal approaches are superior to unimodal ones, as important information relevant to music understanding can be embedded in different modalities and the interplay between them. Additionally, we developed a global aggregation methodology that assists in the interpretation of the correlation between genres or emotions and their associated audio and lyrical features, offering a comprehensive overview of the most representative characteristics of each class.
% We evaluated the robustness of our methodology through its application to two distinct datasets and two different tasks. 

Future research will focus on enhancing \textsc{MusicLIME} by improving various pipeline components, including data preprocessing, encoding techniques, and strategies for sample selection and perturbation within the core \textsc{LIME} algorithm. To address the current challenge of defining clear criteria for evaluating the quality of generated explanations, we plan to conduct a human evaluation survey to assess their effectiveness in enhancing music understanding and interpretation. Additionally, since the lyrical modality is currently analyzed at the word level, which may overlook broader contextual meaning, we aim to make \textsc{MusicLIME} more context-aware  to capture more general ideas beyond individual words. Additionally, we will investigate alternative explanation methods, such as counterfactual explanations, and assess their applicability in a multimodal framework for music understanding.

\section*{Acknowledgments}

We express our sincere gratitude to Christos Garoufis for his useful
insights and thoughtful comments on our work. The research project is implemented in the framework of H.F.R.I call “Basic research Financing (Horizontal support of all Sciences)” under the National Recovery and Resilience Plan “Greece 2.0” funded by the European Union –NextGenerationEU(H.F.R.I. Project Number: 15111 - Emotional Artificial Intelligence in Music Expression).

\bibliographystyle{IEEEbib}
\bibliography{strings,refs}
\end{document}